\documentclass[12pt]{article}

\newcommand{\x}{arXiv:}
\newcommand{\m}{\mathrm}
\newcommand{\be}{\begin{equation}}
\newcommand{\ee}{\end{equation}}
\newcommand{\ba}{\begin{eqnarray}}
\newcommand{\ea}{\end{eqnarray}}

\usepackage{graphicx}
\usepackage{amssymb}
\usepackage{amsmath}
\usepackage[T1]{fontenc} 
\usepackage[ansinew]{inputenc} 
\usepackage[nosort]{cite}
\newcommand{\inbar}{\vrule height1.57ex width.4pt depth0pt}
\newcommand{\SW}{\relax{\hbox{$\ \inbar\kern-.285em{\rm S}$}}}

\begin{document}
\thispagestyle{empty}
\begin{center}

\null \vskip-1truecm \vskip2truecm

{\Large{\bf \textsf{Extremal Bifurcations of Rotating AdS$_4$ Black Holes}}}

{\large{\bf \textsf{}}}

{\large{\bf \textsf{}}}

\vskip1truecm

{\large \textsf{Brett McInnes}}

\vskip1truecm

\textsf{\\  National
  University of Singapore}

\textsf{email: matmcinn@nus.edu.sg}\\

\end{center}
\vskip1truecm \centerline{\textsf{ABSTRACT}} \baselineskip=15pt
\medskip

The Weak Gravity Conjecture arises from the assertion that all extremal black holes, even those which are ``classical'' in the sense of being very massive, must decay by quantum-mechanical emission of particles or smaller black holes. This is interesting, because some observed astrophysical black holes are on the brink of being extremal ---$\,$ though this is due to rapid rotation rather than a large electric or magnetic charge. The possibility that rotating near-extremal black holes might, in addition to radiating spinning particles, also bifurcate by emitting smaller black holes, has attracted much attention of late. There is, however, a basic question to be answered here: can such a bifurcation be compatible with the second law of thermodynamics? This is by no means clear. Here we show that, if there is indeed such a mechanism for bifurcations of AdS$_4$-Kerr-Newman black holes, then this process can in fact satisfy the second law.

\newpage

\addtocounter{section}{1}
\section* {\large{\textsf{1. Bifurcations of Black Holes }}}
Rotating extremal black holes are of great interest observationally, because some observed astrophysical black holes are on the brink of extremality. For example, it has recently been claimed \cite{kn:cygnus} that Cygnus X-1 contains a black hole with dimensionless spin parameter at least $0.9696,$ and possibly as high as $0.9985,$ where unity represents extremality. The more speculative possibility of observing a black hole which is near-extremal due to a high (magnetic) charge has also attracted much attention recently \cite{kn:juan1,kn:yang,kn:ullah}.

These objects are also of the utmost theoretical interest, for many reasons. Particularly relevant to the astrophysical case, which of course involves classical black holes, is the following statement: \emph{extremal black holes decay due to quantum effects, even when they are very massive}, so that one might have expected them to be ``completely classical'', since they do not emit Hawking radiation. This statement is a consequence of the idea on which the well-known ``Weak Gravity Conjecture'', or WGC \cite{kn:motl,kn:palti,kn:rude}, is founded: the claim that extremal Reissner-Nordstr\"{o}m black holes ---$\,$ but the same motivations apply to the Kerr-Newman case ---$\,$ must \emph{always} (special, symmetry-protected exceptions apart) decay through quantum-mechanical emission of particles and/or smaller black holes.

The decay channel through the emission of particles can be expected to dominate over the process of emission of black holes, since particles will normally have the higher charge/mass ratio. However, the latter presumably also occurs, at some unknown rate. Even though the rate is much lower than for particles, this process could well have a very different observational signature, and so might still be discernible. Even if that proves not to be correct, the mere assertion that it is \emph{possible} for an extremal black hole to split in this manner has profound ramifications for our general understanding of quantum gravity, as was pointed out in \cite{kn:motl,kn:kats}: see for example the recent work of Arkani-Hamed et al. \cite{kn:NAH} on the consequences for the charge/mass ratio of extremal black holes. Notice that the emission of one black hole by another is indeed a probe of \emph{quantum} gravity, because four-dimensional classical black holes can never behave in this manner: see page 308 of \cite{kn:wald}.

In order to distinguish the two decay channels, we will refer to the emission of a black hole by a larger extremal black hole as a quantum \emph{extremal bifurcation} of the original black hole. In this work, we are concerned exclusively with such quantum extremal bifurcations, though we certainly do \emph{not} wish to suggest that the alternative, ``particle'' channel is not important.

There are two key questions to be answered before we can study such bifurcations. The first is: can extremal bifurcation be consistent with the basic laws of black hole thermodynamics \cite{kn:bekhod}, which should be respected even by quantum black holes? \emph{This is very far from obvious.} If we can answer affirmatively, we can work towards an answer to the second question: is the \emph{rate} at which this exotic process takes place sufficiently large as to be astrophysically interesting?

Here we hope to shed some light on these questions by studying the analogous problem for asymptotically anti-de Sitter, four-dimensional\footnote{We do not know how to extend our results to higher dimensions, because of the remarkable fact that the Kerr-Newman metric has yet to be generalized to dimensions higher than four, except for approximations valid only in special cases: see \cite{kn:emparan}.} Kerr-Newman black holes. This strategy is motivated partly by the hope that the lessons we will learn might carry over in some form to the asymptotically flat case, partly by the possibility of exploiting the AdS/CFT duality \cite{kn:casa,kn:nat,kn:bag} (though, for the most part, we leave that to future work). Note that the WGC has been studied extensively in the asymptotically AdS context: see for example \cite{kn:qing,kn:naka1,kn:mig,kn:crem,kn:agar,kn:naka2,kn:102}.

In order to proceed, one must be confident that the products of extremal bifurcation are indeed black holes, that is, that they have event horizons, with which it is possible to associate well-defined entropies. In other words, one needs Cosmic Censorship to hold. There are good reasons to believe that this is indeed so in the AdS case: in fact, this is itself closely associated with the WGC; see for example \cite{kn:weak,kn:horsant,kn:suvrat,kn:bala,kn:sean,kn:wang,kn:netta,kn:boge}. We will therefore assume in this work that Censorship does hold in the AdS context at least.

The most basic problem we must deal with is simply this: the mere existence of a mechanism for some process to occur does not mean that it \emph{will} occur. To take a relevant example: rotating black holes in high spacetime dimensions are capable of dividing into two smaller black holes, due to centrifugal ``forces'', in a manner reminiscent of extremal bifurcation \cite{kn:empmy} (see also \cite{kn:pau,kn:empa}). But although this ``decay channel'' is always available, such a fission does not of course always occur ---$\,$ a sufficiently high angular momentum is required. Emparan and Myers were able to locate this threshold, \emph{without} analysing the extremely complex fission process itself, by determining the circumstances under which the total entropy of the putative final pair of black holes exceeded that of the original black hole. In short: given that a mechanism (which may not be fully understood) for some process exists, the second law of black hole thermodynamics allows us to determine whether that process \emph{actually proceeds}.

In the same way, the existence of a hypothetical ``extremal bifurcation'' channel for extremal black holes does not ensure that these objects will actually decay in this manner. However, a thermodynamic argument might settle that question, even if the physics of the splitting process itself is not yet fully understood.

If the emitted object is approximately classical, the entropy can be computed by using the Hawking area formula\footnote{There has been some debate (see for example \cite{kn:seanlisa}) as to whether the Hawking formula is valid for extremal black holes. Here we accept the straightforward positive answer provided by string theory \cite{kn:andrew}.}; otherwise it can be done using the well-known quantum-corrected versions of that formula (see for example \cite{kn:pando,kn:xav} and their references). \emph{The hope is that one can prove that the total entropy of the hypothetical final pair exceeds the entropy of the original black hole.}

This could be very useful, in at least two ways. First, of course, it would reassure us that extremal bifurcation is not \emph{obstructed} by the second law ---$\,$ a very real possibility, particularly in four dimensions, where, unlike in higher dimensions, black hole angular momenta are strongly constrained by Cosmic Censorship. Secondly, quantitative results in this direction might ultimately help to settle the all-important question for eventual astrophysical applications: how rapidly does extremal bifurcation occur?

In the pure Reissner-Nordstr\"{o}m case, an argument of this kind has been given in \cite{kn:hod} (see also \cite{kn:urban} for the asymptotically AdS case). That is, given an extremal Reissner-Nordstr\"{o}m black hole, one can use the second law\footnote{The entropies are not actually calculated in \cite{kn:hod}. Instead, an ingenious argument, which involves applying the ``universal relaxation bound'' \cite{kn:hodrelax} to the perturbation modes of extremal Reissner-Nordstr\"{o}m black holes, is used; this bound follows from the generalised second law. Unfortunately it is not clear that this argument can be extended to the much more complicated Kerr-Newman geometry.} to show that, if there indeed exists such a decay channel for this object, a small quantum black hole will be emitted. (It is recognised as such by the fact that it violates classical Cosmic Censorship, which is replaced by a suitable quantum-corrected version \cite{kn:kats}.) The question now is whether a result of this general kind can be established for the \emph{rotating} case, that is, for Kerr-Newman and AdS$_4$-Kerr-Newman black holes.

A thermodynamic approach to understanding extremal bifurcation is particularly natural in view of recent work \cite{kn:rem1,kn:rem2,kn:goon} which relates the ``quantum'' version of Censorship demanded by the WGC to the \emph{positive} shift in the black hole entropy implied by taking non-negligible higher-dimension operators into account. In short, the entropy of quantum-corrected black holes is ``larger than expected'', and of course this is welcome if one hopes to show, using the second law, that extremal bifurcation actually takes place\footnote{The calculations in \cite{kn:rem1,kn:rem2,kn:goon} are concerned exclusively with (AdS)-Reissner-Nordstr\"{o}m black holes. However, the entropy shift is positive also when the emitted object is AdS$_4$-Kerr and has a smaller mass than classical Censorship allows in that case: see \cite{kn:aalsma}.}.

However, while it is certainly important and encouraging that the entropy shift computed in \cite{kn:rem1,kn:rem2,kn:goon,kn:aalsma} is ``in the right direction\footnote{See however \cite{kn:cano}; work remains to be done before one can claim that the WGC is fully understood in the string-theoretic embedding, particularly for rotating black holes.},'' the actual value of the shift depends on Wilson coefficients which are known only in principle. Therefore, even if we can construct an expression for the entropy of the quantum-corrected emitted black hole, as in \cite{kn:pando,kn:xav}, we will not be able to extract an explicit numerical value for it. It is therefore not clear that we can show that the entropy of the emitted object is large enough to (over-)compensate for the decrease in the entropy of the original black hole after it has emitted a smaller one.

It begins to seem that it may not be possible, in practice, to show that ``rotational extremal bifurcation'' follows, even for classical rotating black holes, from the second law. However, there is a loophole here.

It is true that, if an extremal AdS$_4$-Kerr-Newman black hole decays by emitting a small charged, non-rotating black hole, then the latter must have a charge-to-mass ratio which violates classical Censorship (so that gravity is still the weakest ``force'' when AdS$_4$-Kerr-Newman black holes replace Reissner-Nordstr\"{o}m black holes \cite{kn:104}). That is, the emitted black hole must in this case be a quantum-corrected black hole, with an entropy that is difficult to estimate quantitatively.

However, if the putative decay is by emission of a small, uncharged, \emph{rotating} black hole, then the emitted object can, in some cases, satisfy \emph{classical} Cosmic Censorship \cite{kn:104}. This means that its entropy can be computed in the usual manner. We are therefore in a position to answer the following modified version of our original question: given \emph{any} extremal rotating AdS$_4$-Kerr-Newman black hole, is it possible to find a pair of rotating black holes \emph{satisfying classical Censorship} which might, consistently with the conservation laws and the second law of thermodynamics, be the result of the decay of that black hole? In other words, does the emitted black hole in \emph{this} case always have sufficient entropy to over-compensate for the decrease in the entropy of the original black hole?

We will show that this is always so. It follows that, since including the effect of higher-dimension operators increases the entropy \cite{kn:rem1,kn:rem2,kn:goon,kn:aalsma} relative to classical expectations, the second law ensures that, if the conjectured decay channel for the original black hole exists, that black hole will \emph{always} decay through the emission of a quantum-corrected black hole. This is so even when the original black hole is so massive as to be ``classical''.

The surviving black hole, if it is extremal, will likewise decay, and the process will continue, decreasing the angular momentum at each stage, until the black hole is either essentially Reissner-Nordstr\"{o}m or no longer classical. In the Reissner-Nordstr\"{o}m case, the methods of \cite{kn:hod} then complete the argument.

We begin with a r\'{e}sum\'{e} of the main results of \cite{kn:104}, and a discussion of the entropy of AdS$_4$-Kerr-Newman black holes.

\addtocounter{section}{1}
\section* {\large{\textsf{2. Relevant Properties of AdS$_4$-Kerr-Newman Black Holes}}}
\addtocounter{section}{1}
\subsection* {\large{\textsf{2.1. Cosmic Censorship for AdS$_4$-Kerr-Newman Black Holes}}}
We will discuss \emph{magnetically} charged black holes \cite{kn:weinberg}, partly because such objects are currently of observational interest \cite{kn:yang,kn:ullah}, but mainly because they are \emph{generically} close to extremality \cite{kn:juan1}, so that extremal bifurcation is relevant to them. However, the reader who is interested in the electric or dyonic cases can replace the magnetic parameter $P$ throughout our discussion with $\sqrt{P^2 + Q^2}$, where $Q$ is the electric charge parameter.

The geometry we wish to study is described by a metric of the form \cite{kn:cognola}
\begin{flalign}\label{A}
g(\m{AdS}_{a,M,P,L}) = &- {\Delta_r \over \rho^2}\Bigg[\,\m{d}t \; - \; {a \over \Xi}\sin^2\theta \,\m{d}\phi\Bigg]^2\;+\;{\rho^2 \over \Delta_r}\m{d}r^2\;+\;{\rho^2 \over \Delta_{\theta}}\m{d}\theta^2 \\ \notag \,\,\,\,&+\;{\sin^2\theta \,\Delta_{\theta} \over \rho^2}\Bigg[a\,\m{d}t \; - \;{r^2\,+\,a^2 \over \Xi}\,\m{d}\phi\Bigg]^2,
\end{flalign}
where
\begin{eqnarray}\label{B}
\rho^2& = & r^2\,+\,a^2\cos^2\theta, \nonumber\\
\Delta_r & = & (r^2+a^2)\Big(1 + {r^2\over L^2}\Big) - 2Mr + {P^2\over 4\pi},\nonumber\\
\Delta_{\theta}& = & 1 - {a^2\over L^2} \, \cos^2\theta, \nonumber\\
\Xi & = & 1 - {a^2\over L^2}.
\end{eqnarray}
Here $L$ is the asymptotic AdS$_4$ length scale, $a$ is the specific angular momentum (angular momentum per unit physical mass, with units of length), and $M$ and $P$ are parameters with units of length which, together with $a$, determine \cite{kn:gibperry} the physical mass $\mathcal{M}$ and the physical magnetic charge $\mathcal{P}$ through
\begin{equation}\label{C}
\mathcal{M}\;=\;M/(\ell_{\textsf{G}}^2\Xi^2), \;\;\;\;\;\mathcal{P}\;=P/(\ell_{\textsf{G}}\Xi),
\end{equation}
where $\ell_{\textsf{G}}$ is the AdS$_4$ gravitational length scale. (We use natural (``particle physics'') units, in which neither $L$ nor $\ell_{\textsf{G}}$ is equal to unity.)

The outer horizon is located at $r = r_{\textsf{H}}$, which is related to the parameters in the usual manner:
\begin{equation}\label{D}
\Delta_r(r_{\textsf{H}})\;=\;(r_{\textsf{H}}^2+a^2)\Big(1 + {r_{\textsf{H}}^2\over L^2}\Big) - 2Mr_{\textsf{H}} + {P^2\over 4\pi}\;=\;0.
\end{equation}
This equation has a relatively simple solution in the extremal case:
\begin{equation}\label{E}
r_{\textsf{H\;ext}}^2 \;=\; {L^2\over 6}\left(-1 - {a^2\over L^2} + \sqrt{\left(1 + {a^2\over L^2}\right)^2 + {12\over L^2}\left(a^2\,+\,{P^2\over 4\pi}\right)}\right).
\end{equation}
\begin{figure}[!h]
\centering
\includegraphics[width=0.7\textwidth]{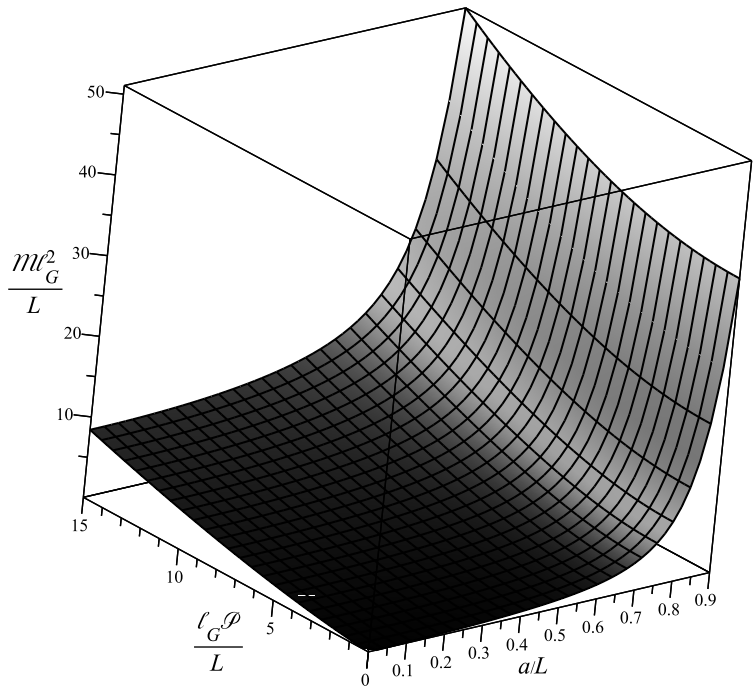}
\caption{AdS$_4$-Kerr-Newman black holes with $a/L < 1$ satisfying classical Cosmic Censorship correspond to points on or above the surface shown.}
\end{figure}
\begin{figure}[!h]
\centering
\includegraphics[width=0.7\textwidth]{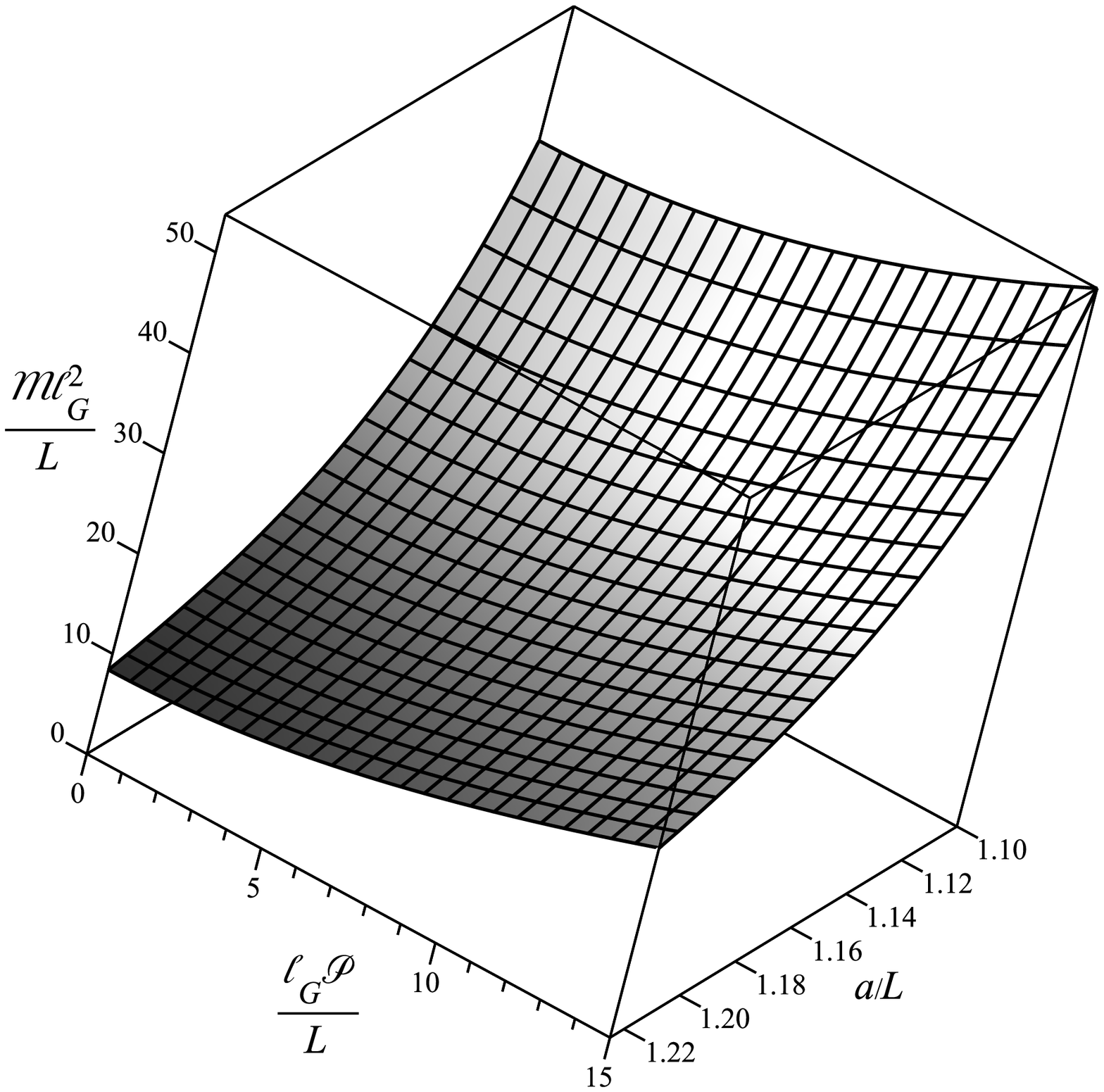}
\caption{AdS$_4$-Kerr-Newman black holes with $a/L > 1$ satisfying classical Cosmic Censorship correspond to points on or above the surface shown.}
\end{figure}
Using this, together with the relations (\ref{C}), one arrives at the condition for classical Cosmic Censorship to hold:
\begin{equation}\label{F}
\mathcal{M} \;\geq \; \Theta\left(a/L,\,\mathcal{P}\right)\;\equiv \;{\sqrt{6}L \left[{1\over 36} \left({5a^2\over L^2}\, -\, 1\, + \textsf{Z} \;\right) \times \left(5\, -\, {a^2\over L^2}\, +\, \textsf{Z} \;\right) + \,{\ell_{\textsf{G}}^2\left(1 - [a^2/L^2]\right)^2\mathcal{P}^2\over 4\pi L^2} \;\right]\over 2\ell_{\textsf{G}}^2\left(1 - [a^2/L^2]\right)^2\sqrt{-1\, -\, {a^2\over L^2} \,+\,\textsf{Z} }}
\end{equation}
where $\Theta\left(a/L,\,\mathcal{P}\right)$ is the function of dimensionless quantities defined as shown, and where
\begin{equation}\label{G}
\textsf{Z}\;=\;\sqrt{\left(1 + {a^2\over L^2}\right)^2 + {12\over L^2}\left(a^2\,+\,{\ell_{\textsf{G}}^2\left(1 - [a^2/L^2]\right)^2\mathcal{P}^2\over 4\pi}\right)}.
\end{equation}

The function $\Theta\left(a/L,\,\mathcal{P}\right)$ is graphed in Figures\footnote{Notice that, in these Figures and throughout our discussion, $\mathcal{M}$ and $\mathcal{P}$ occur in combination with the gravitational length scale $\ell_{\textsf{G}}$. In the AdS/CFT correspondence \cite{kn:casa,kn:nat,kn:bag} this quantity can be regarded as a fixed multiple of $L$: from \cite{kn:ABJM,kn:AdS4} we have  $${L\over\ell_{\textsf{G}}} = {\left(2N_{\textsf{c}}\right)^{3/4}\over \sqrt{3}},$$
where $N_{\textsf{c}}$ is the number of coincident M2-branes in the dual description; this number is normally assumed to be very large.} 1 and 2. One sees that it is possible to satisfy Censorship for values of $a/L$ which are either below unity \emph{or above it}, provided that, in both cases, $a/L$ is not too close to unity. For example, if we set $\mathcal{M}\ell_{\textsf{G}}^2/L = 20, \; \mathcal{P}\ell_{\textsf{G}}/L = 15,$ a numerical computation using (\ref{F}) shows that the acceptable values of $a/L$ satisfy \emph{either} $a/L \leq \approx 0.788$ \emph{or} $a/L \geq \approx 1.214$ (and this can be seen in the Figures).

In short, Censorship excludes only a \emph{band} of values for $a/L$ around unity: it does \emph{not} exclude large values of $a/L$. The values of $a/L$ compatible with Censorship fall into two disjoint subsets, and the corresponding black holes cannot ``communicate'' by any continuous process. They might, however, be related by a discontinuous, quantum process, and in fact that is precisely what occurs in the course of extremal bifurcation, as we shall see.

The ``\emph{exotic}'' black holes with $a/L > 1$ have a very unusual geometric structure; see \cite{kn:104} for the details. On the other hand, however, arbitrarily demanding (say) $a/L \leq k$ for some positive constant $k < 1$ would mean that one is imposing a \emph{lower} bound on the masses of emitted black holes with given angular momenta, because $a$ is the ratio of the angular momentum to the mass. This in itself does not seem very reasonable; intuitively, it would seem that objects with small masses should be the easiest to create by a quantum process. Furthermore, any such lower bound would be fixed by the \emph{asymptotic} curvature scale $L$, which does not seem to be relevant to the local physics. From a physical point of view, then, we can say that any proposed bounds on $a/L$, apart from those due to Cosmic Censorship, would have to be supported by a strong argument.

We will return to this later; for the present we simply note that allowing $a > L$ does not cause any difficulties for the calculations we need to perform here.

Next we turn to the question of the classical stability of these black holes.

\addtocounter{section}{1}
\subsection* {\large{\textsf{2.2. Superradiant Modes for AdS$_4$-Kerr-Newman Black Holes}}}
It has long been known \cite{kn:reall} that all extremal AdS-Kerr black holes are classically unstable to the emission of superradiant modes \cite{kn:super}. (For recent results on the superradiant instability of AdS$_4$-Kerr black holes, see \cite{kn:ches}.) However, this is not always the case for extremal AdS$_4$-Kerr-Newman black holes. When $a/L < 1,$ they can be stable if their electric or magnetic charge is sufficiently large \cite{kn:104}, exceeding a certain minimal value $\mathcal{P}_{\textsf{min}}$, which of course depends on $a$:
\begin{equation}\label{H}
\mathcal{P} \;\geq\;\mathcal{P}_{\textsf{min}}\;=\;{2\,\sqrt{\pi a L}\over \ell_{\textsf{G}}\left(1-{a\over L}\right)}.
\end{equation}
When $a/L > 1,$ by contrast, extremal AdS$_4$-Kerr-Newman black holes are stable against superradiance for \emph{all} values of the charge, including zero \cite{kn:104}.

In order to be able to focus exclusively on the extremal bifurcation process, we shall only consider black holes which are classically stable: that is, they either satisfy (\ref{H}) when $a/L < 1,$ or they have $a/L > 1.$

\addtocounter{section}{1}
\subsection* {\large{\textsf{2.3. Entropy of Extremal AdS$_4$-Kerr-Newman Black Holes}}}
\begin{figure}[!h]
\centering
\includegraphics[width=0.5\textwidth]{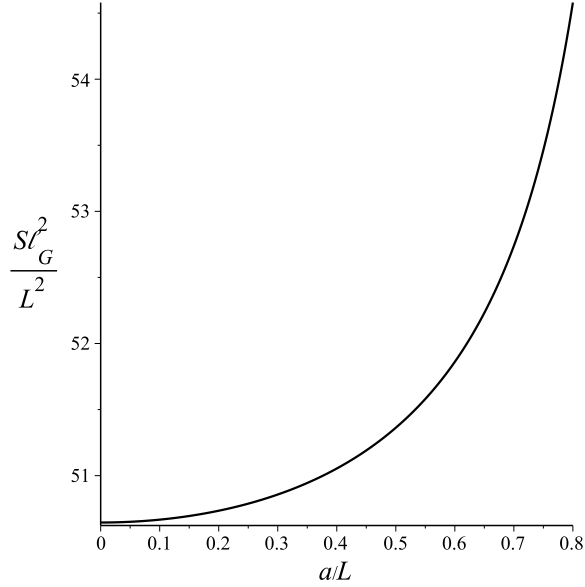}
\caption{The entropy (scaled by the dimensionless factor $\ell^2_{\textsf{G}}/L^2$) of extremal AdS$_4$-Kerr-Newman black holes with $a/L < 1$ for $\ell_{\textsf{G}}\mathcal{P}/L = 100$, as a function of $a/L$.}
\end{figure}
\begin{figure}[!h]
\centering
\includegraphics[width=0.5\textwidth]{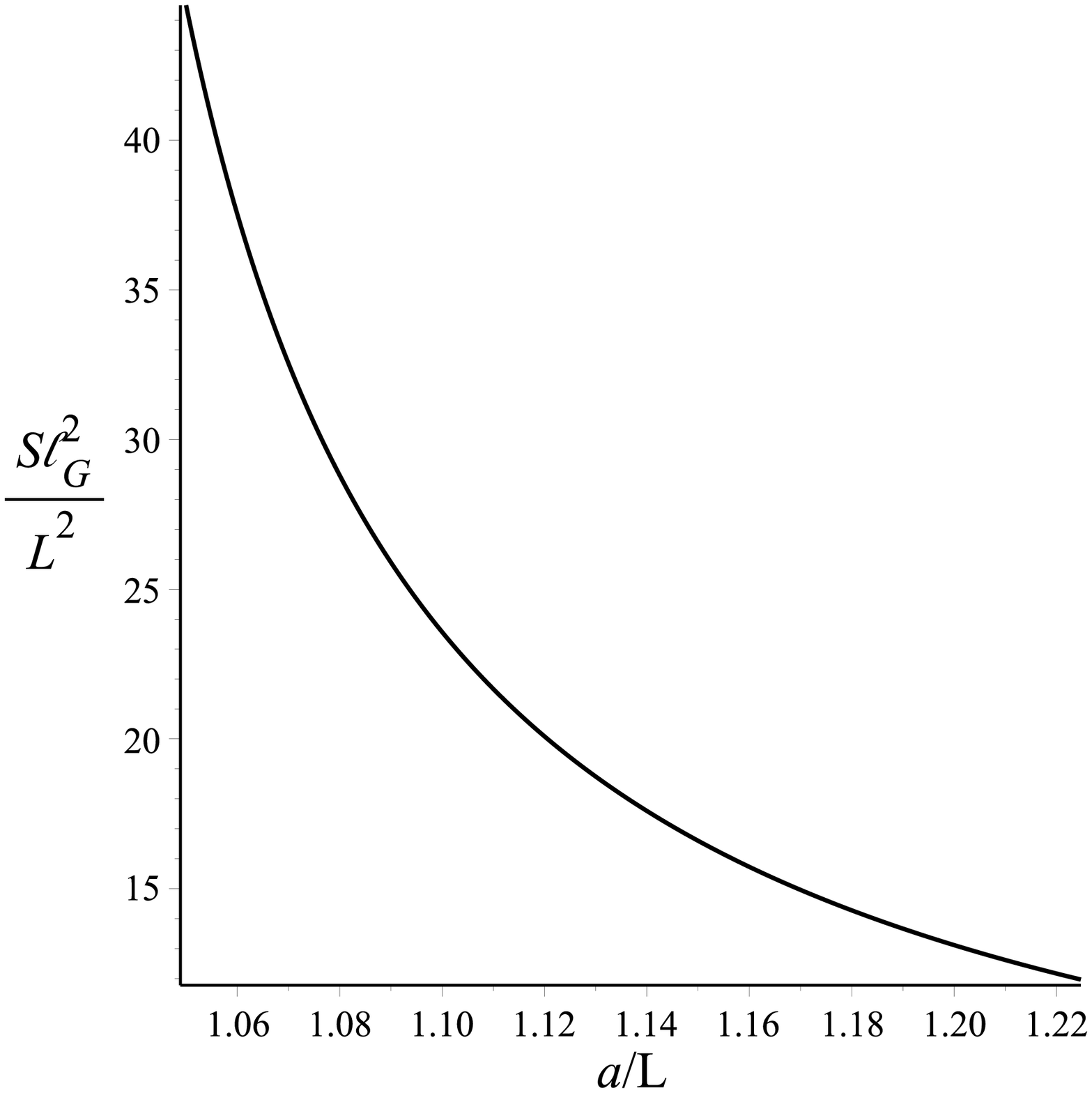}
\caption{The entropy (scaled by the dimensionless factor $\ell^2_{\textsf{G}}/L^2$) of extremal AdS$_4$-Kerr-Newman black holes with $a/L > 1$ for $\ell_{\textsf{G}}\mathcal{P}/L = 0$, as a function of $a/L$.}
\end{figure}
\begin{figure}[!h]
\centering
\includegraphics[width=0.5\textwidth]{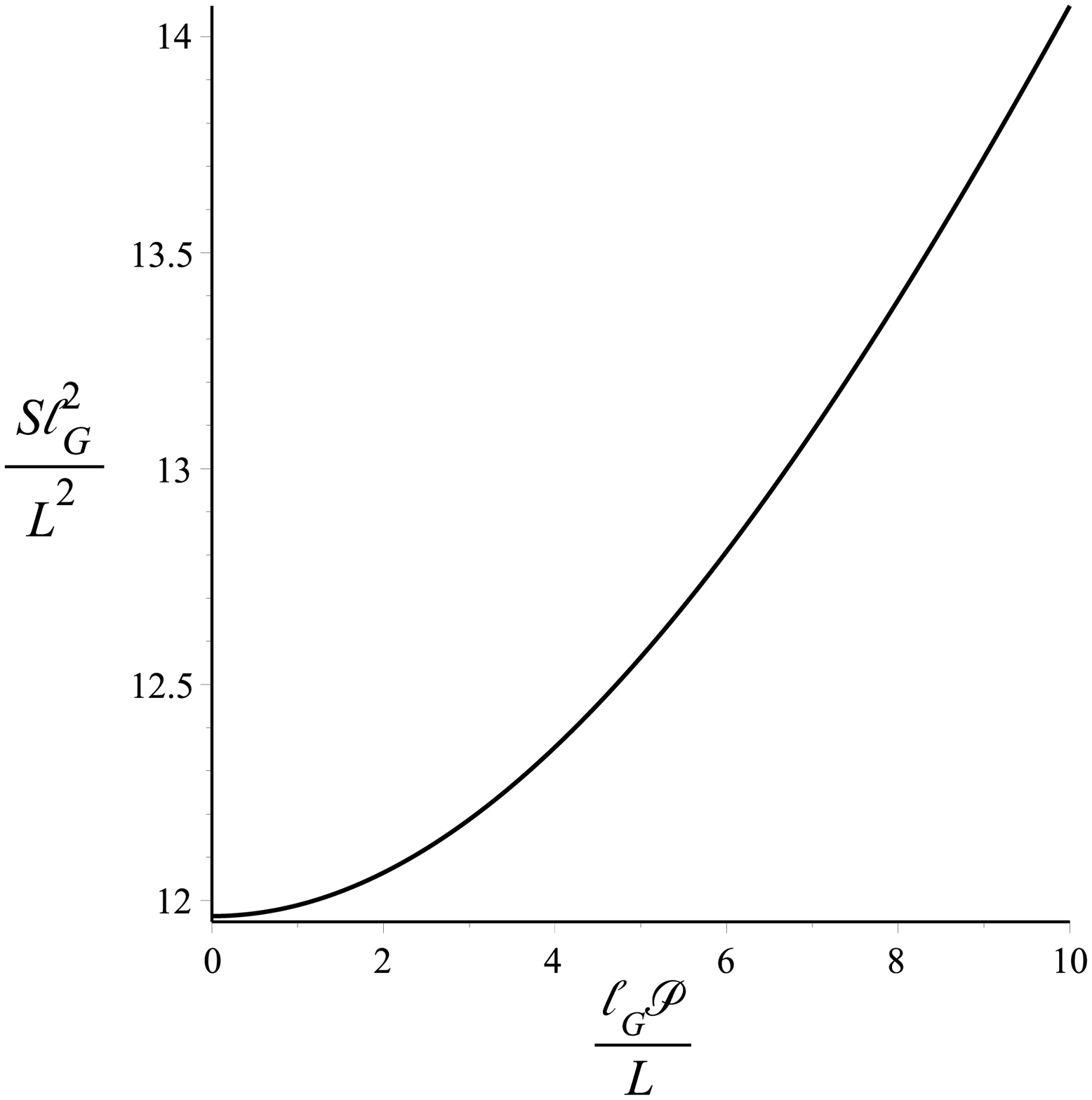}
\caption{The entropy (scaled by the dimensionless factor $\ell^2_{\textsf{G}}/L^2$) of extremal AdS$_4$-Kerr-Newman black holes, with $a/L =\sqrt{3/2}$, as a function of  $\ell_{\textsf{G}}\mathcal{P}/L$.}
\end{figure}
The entropy of any AdS$_4$-Kerr-Newman black hole is computed from the area of the two-spheres with $t = $ constant, $r = r_{\textsf{H}}:$
\begin{equation}\label{J}
S\;=\;{\pi \,\left(r_{\textsf{H}}^2 + a^2\right)\over \ell_{\textsf{G}}^2|1 - (a/L)^2|}.
\end{equation}
Using equations (\ref{C}) and (\ref{E}) we can now evaluate the entropy of an extremal AdS$_4$-Kerr-Newman black hole:
\begin{equation}\label{K}
S_{\textsf{ext}}\;=\;{\pi \,L^2\left(-1 + {5a^2\over L^2} + \sqrt{\left(1 + {a^2\over L^2}\right)^2 + {12\over L^2}\left(a^2\,+\,{\ell_{\textsf{G}}^2\left(1 - [a^2/L^2]\right)^2\mathcal{P}^2\over 4\pi}\right)}\right)\over 6\ell_{\textsf{G}}^2|1 - (a/L)^2|}.
\end{equation}

We can think of $a$ and $\mathcal{P}$ as independent variables here, with the understanding that the mass of the black hole can be adjusted so that it remains extremal.
Figures 3, 4, and 5 show the dependence of $S_{\textsf{ext}}$ on $a/L$ and $\ell_{\textsf{G}}\mathcal{P}/L$ for certain fixed values of the other parameter, chosen for reasons to be explained. We see that the entropy increases with $a/L$ (for fixed $\mathcal{P}$) when $a/L < 1,$ but decreases when $a/L > 1;$ on the other hand, it increases with $\mathcal{P}$ (for fixed $a/L$) when $a/L > 1.$

\addtocounter{section}{1}
\subsection* {\large{\textsf{2.4. Extremal Bifurcation of AdS$_4$-Kerr-Newman Black Holes}}}
The physics of asymptotically AdS spacetimes is generically periodic in time: for example, all timelike geodesics, that is, with all possible timelike initial tangent vectors, emanating from an event in AdS itself intersect periodically, with period $2\pi L$, where $L$ is the curvature scale \cite{kn:orb}. If $L$ is very large, then, the period will also be very large.

If we take $L$ to be large, and suppose that an AdS$_4$-Kerr-Newman black hole decays into two black holes which move apart at relativistic speeds ---$\,$ we will have to verify this ---$\,$ then we can assume that an approximate local thermodynamic equilibrium is established long before the effects of the periodicity make themselves felt, and so a thermodynamic approach to our problem is appropriate. Understanding the dissipative physics responsible for the irreversibility here would require a detailed description of the mechanism of extremal bifurcation, but perhaps the methods of \cite{kn:empa} might be applicable to some extent.

Now in fact, in applications of gauge-gravity duality, it is almost invariably assumed that $L$ \emph{is} indeed much larger than any other length scale in the system under consideration \cite{kn:casa,kn:nat,kn:bag}. Thus, the standard assumptions of applied holography automatically ensure that the ``separation of relaxation times'' \cite{kn:dover}, needed for a thermodynamic approach to the problem to be appropriate, does actually hold.

We begin with an extremal AdS$_4$-Kerr-Newman black hole, with mass $\mathcal{M},$ magnetic charge $\mathcal{P}$, and angular momentum $\mathcal{J}$, and set $\mathcal{A} = \mathcal{J}/\mathcal{M}.$ We assume that this black hole is sufficiently massive that, to a good approximation, it is ``classical": in particular, it satisfies classical Cosmic Censorship. We also assume that it is not ``exotic'', that is, $\mathcal{A}/L < 1.$ We will take it that the black hole is sufficiently highly charged that the stability condition (\ref{H}) is satisfied.

We will now consider the possibility that this black hole emits a small black hole with mass, charge, and angular momentum $m, p, j,$ which we take to be all positive or zero. We set $a = j/m.$

After the bifurcation, the parameters of the new large black hole (let us call it the ``survivor'') are $\mathcal{M} + \delta\mathcal{M}, \mathcal{P} + \delta\mathcal{P}, \mathcal{J} + \delta\mathcal{J}.$ We assume as usual that the survivor is still so massive that it satisfies classical Censorship, that it is still so highly charged that (\ref{H}) continues to hold, and that the emitted object is moving relative to the survivor, so that $m \;<\; -\,\delta\mathcal{M}$.

We assume further that the bifurcation can occur \emph{either} through emission of a charged, non-rotating black hole, \emph{or} by emission of a neutral, rotating (AdS$_4$-Kerr) black hole.

In the former case, one obtains \cite{kn:104}, using charge conservation, the familiar inequality
\begin{equation}\label{L}
{p \over m} \;>\; 2\sqrt{\pi}\ell_{\textsf{G}};
\end{equation}
that is, the emitted object necessarily violates classical Censorship for AdS$_4$-Reissner-Nordstr\"{o}m black holes. This is the usual consequence of extremal bifurcation in the non-rotating case.

Turning to the more complex case in which an AdS$_4$-Kerr black hole is emitted ---$\,$ this is the only case we consider henceforth ---$\,$ let us focus on the situation in which this process does not destroy the axial symmetry of the system, so that the axial Killing vector survives and we can use conservation of the corresponding angular momentum; that is, we study the case where the emitted black hole moves away in the direction of the axis of symmetry. Then $j \;=\;-\,\delta\mathcal{J},$ and using this one finds \cite{kn:104} that $a/L,$ regarded as a function of the parameters of the original black hole, must satisfy
\begin{equation}\label{M}
{a\over L} \;>\;\textsf{K}\left(\mathcal{P},\;\mathcal{A}\right),
\end{equation}
where the dimensionless quantity $\textsf{K}\left(\mathcal{P},\;\mathcal{A}\right)$ is a known but very complicated function\footnote{This function can, after considerable simplification, can be expressed as follows. To avoid being overwhelmed by notation, we define dimensionless parameters $\alpha \equiv \mathcal{A}/L,$ and $\varpi \equiv \ell_{\textsf{G}}\mathcal{P}/L.$ Then expressing $\textsf{K}\left(\mathcal{P},\;\mathcal{A}\right)$ as $\textsf{K}\left(\varpi,\;\alpha\right),$ we have
\begin{equation}\label{EXPLICITK}
\textsf{K}\left(\varpi,\;\alpha\right) \;=\;\textsf{N}\left(\varpi,\;\alpha\right)/\textsf{D}\left(\varpi,\;\alpha\right),
\end{equation}
where
\begin{eqnarray}\label{VERYLONGN}
\textsf{N}\left(\varpi,\;\alpha\right)& = &(-4\alpha^6+66\alpha^4+2)\pi^{3/2}+18\alpha^2\varpi^2\sqrt{\pi}(\alpha^2-1)^2)\sqrt{(3\varpi^2+\pi)\alpha^4+(-6\varpi^2 +14\pi)\alpha^2+3\varpi^2+\pi}   \nonumber\\
 &  & +(9((\varpi^2+(4/3)\pi)\alpha^4+(-2\varpi^2+(14/3)\pi)\alpha^2+\varpi^2-(2/3)\pi))((\varpi^2+(1/3)\pi)\alpha^4\nonumber\\
&  &  +(-2\varpi^2+(14/3)\pi)\alpha^2+\varpi^2+(1/3)\pi, \nonumber\\
\end{eqnarray}
and
\begin{eqnarray}\label{VERYLONGD}
\textsf{D}\left(\varpi,\;\alpha\right)& = & 9\alpha(((22/3)\alpha^2+(2/9)\alpha^6-4/9)\pi^{3/2}  \nonumber\\
&  &   +2\varpi^2\sqrt{\pi}(\alpha^2-1)^2)\sqrt{(3\varpi^2+\pi)\alpha^4+(-6\varpi^2+14\pi)\alpha^2+3\varpi^2+\pi}                                                                                                                                              \nonumber\\
 &  & +((\varpi^2-(2/3)\pi)\alpha^4+(-2\varpi^2+(14/3)\pi)\alpha^2+\varpi^2+(4/3)\pi)((\varpi^2+(1/3)\pi)\alpha^4\nonumber\\
&  &  +(-2\varpi^2+(14/3)\pi)\alpha^2+\varpi^2+(1/3)\pi). \nonumber\\
\end{eqnarray}
} of the indicated parameters (see \cite{kn:104}). In summary, (\ref{M}) is a necessary condition for angular momentum conservation to be satisfied.

When the original black hole is stable against emission of superradiant modes, that is, when $\mathcal{P}$ and $\mathcal{A}$ satisfy the inequality (\ref{H}) above, one can show \cite{kn:104} that $\textsf{K}(\mathcal{P},\;\mathcal{A}) \geq 1,$ and this, with (\ref{M}), implies
\begin{equation}\label{N}
a\;=\;{j\over m}\;>\;L.
\end{equation}
(Notice that $a/L = 1$ is ruled out; this is compatible with the relations (\ref{C}).) Thus we arrive at the very remarkable conclusion that if an otherwise classically stable extremal AdS$_4$-Kerr-Newman black hole emits a small, uncharged, rotating black hole, then the latter \emph{must} be ``exotic'' in the sense discussed earlier ---$\,$ recall that this means that it is also stable against the emission of superradiant modes, despite having no charge. To put it another way: extremal bifurcation in this case \emph{demands} that it be possible for the emitted black hole to satisfy $a/L > 1.$ It is an intrinsically quantum phenomenon that probes a subspace of the AdS$_4$-Kerr-Newman parameter space that is not continuously connected to the familiar subspace with $a/L < 1.$

In one sense, (\ref{N}) is analogous to (\ref{L}): just as (\ref{L}) requires the charge to be large relative to the mass, forcing us to revise the definition of Censorship \cite{kn:kats}, so also (\ref{N}) requires the angular momentum to be larger, relative to the mass, than one would normally allow. The great difference between the two cases, however, is that (\ref{N}) does \emph{not} necessarily entail a violation of classical Cosmic Censorship. We will return to this crucial observation, below. (Notice too that the length scales which must be present for dimensional reasons are very different in (\ref{L}) and (\ref{N}).)

\addtocounter{section}{1}
\section* {\large{\textsf{3. Fixing the Parameters of the Emitted Black Hole }}}
Our objective now is to give an explicit construction of a candidate pair of black holes which might be produced by the fission of an extremal AdS$_4$-Kerr-Newman black hole. We stress that the objective is indeed to produce a \emph{candidate}: as a quantum process, bifurcation can probably occur in many different ways \emph{if} it can occur at all. We do not claim that the channel we investigate is unique or most probable.

The problem is that, in the absence of a precise description of the mechanism underlying extremal bifurcation, it is difficult to specify the parameters of the final pair of black holes in terms of those of the original black hole; but we must do this if we are to compare the final entropy with its initial value. Nevertheless we will argue that this can be done, in the simplest cases, with the aid of some reasonable general assumptions.

Since the original black hole is extremal, its temperature is zero: that is, the system is in its ground state, and so it has no quantum states available to be excited. The simplest procedure is to assume this remains true, at least approximately, if the system bifurcates. That is, both the survivor and the emitted black hole should have temperatures approximately equal to zero, and we will assume this henceforth. If this approximation is for some reason not valid, then the entropies of either the survivor or the emitted black hole, or both, will be larger than our calculations below will indicate: that is, we will be systematically \emph{underestimating} the total final entropy. That would only strengthen our conclusions.

Thus, we are confining our attention to the case in which all three black holes are extremal. This still does not allow us to determine how the energy of the original black hole is distributed between the two final black holes, so we now make the following simple assumptions:

\bigskip

\textbf{[1]}. The value of $a/L$ for the emitted black hole is independent of the charge of the original black hole.

\medskip

\textbf{[2]}. With high probability, the emitted black hole will be created with the largest entropy possible within the set of black holes we are considering, taking into account the constraints imposed by the conservation laws.

\bigskip

Assumption [1] is natural in view of the fact that the emitted black hole is not charged: it is not directly affected by the charge of the original black hole, only very indirectly, through the influence of the charge on the spacetime geometry in the vicinity of the original black hole and the survivor. It seems implausible that this indirect influence should affect the amount of angular momentum to be transferred. Assumption [2] is just a natural consequence of the standard statistical-mechanical probabilistic interpretation of entropy, applied to a system which has been newly created by means of a quantum-mechanical process.

To understand the consequences of these two assumptions, recall that angular momentum conservation implies that the specific angular momentum parameter of the emitted black hole satisfies $a/L > \textsf{K}(\mathcal{P},\;\mathcal{A}),$ where the function $\textsf{K}(\mathcal{P},\;\mathcal{A})$ was introduced in the inequality (\ref{M}). This function is, for fixed $\mathcal{A},$ a monotonically increasing function of $\mathcal{P}$, but it is a \emph{bounded} function: see Figure 6, which shows the special case $\mathcal{A} = 0.2L.$
\begin{figure}[!h]
\centering
\includegraphics[width=0.7\textwidth]{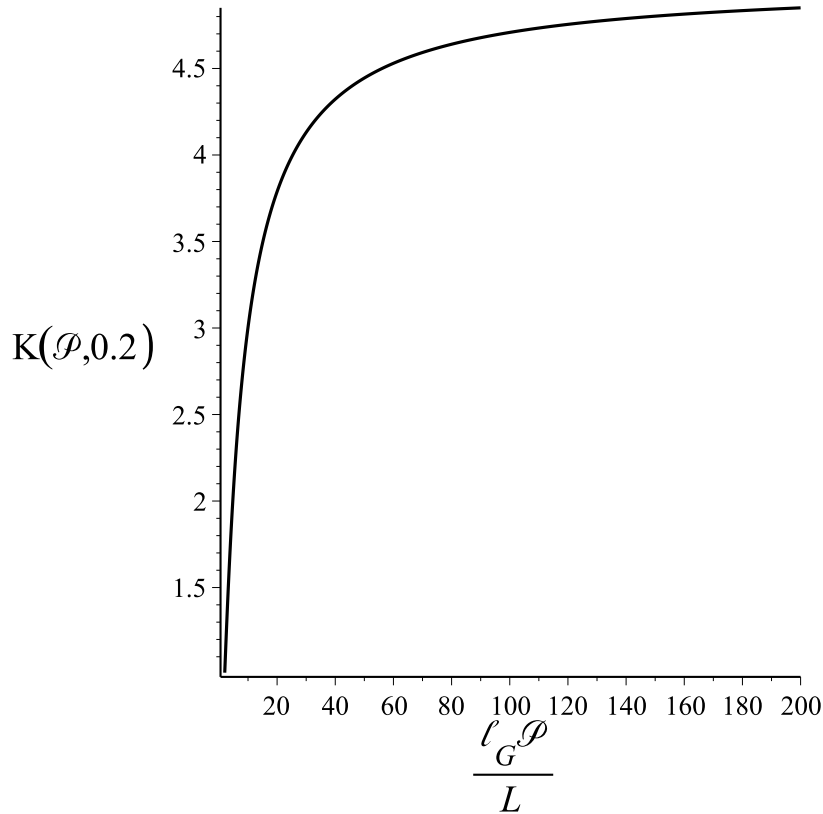}
\caption{The function $\textsf{K}(\mathcal{P},\;\mathcal{A} = 0.2L)$ }
\end{figure}

Since the emitted black hole is uncharged, and $a/L > 1,$ the entropy of the emitted black hole is maximised by taking $a/L$ to be as \emph{small} as angular momentum conservation permits: see Figure 4. Combining this observation with Assumptions [1] and [2], we see that we should take $a/L$ to be the least upper bound of $\textsf{K}(\mathcal{P},\;\mathcal{A})$, regarded as a function of $\mathcal{P}$, with $\mathcal{A}$ fixed.

It is possible to show \cite{kn:104} that this least upper bound, which of course is just the limit of $\textsf{K}(\mathcal{P},\;\mathcal{A})$ as $\mathcal{P} \rightarrow \infty$, is precisely $L/\mathcal{A}$ (so it is equal to 5 in the case shown in Figure 6). Thus we have
\begin{equation}\label{P}
a/L \;=\; L/\mathcal{A}.
\end{equation}
Of course, with this relation, $a/L > 1$ is immediate for these black holes, precisely because $\mathcal{A}/L < 1.$ Even if one were to reject Assumptions [1] and [2], this would clearly be a natural \emph{Ansatz} for $a/L$, since it automatically ensures that the necessary condition for angular momentum conservation, inequality (\ref{M}), is satisfied.

Before we proceed to investigate the consequences of (\ref{P}), we need to confirm that it is compatible with the conservation laws. The conservation of charge is straightforward ---$\,$ all of the original charge is retained by the ``survivor'' ---$\,$ but the conservation of angular momentum is not. Satisfying the condition (\ref{M}) is necessary for it to be possible to satisfy angular momentum conservation, but it is not obvious that it suffices. For that, we need to show that the angular momentum of the putative emitted black hole is smaller than the angular momentum of the original black hole.

The masses of extremal AdS$_4$-Kerr-Newman black holes are given by the function $\Theta\left(a/L,\,\mathcal{P}\right)$ defined in (\ref{F}) above. If we begin with an extremal black hole with a known angular momentum per unit mass $\mathcal{A}$ and a known charge $\mathcal{P},$ then its mass $\mathcal{M}$ is given by $\Theta\left(\mathcal{A}/L,\,\mathcal{P}\right)$. If we fix $a/L$ at the value given in (\ref{P}), then the mass $m$ of the emitted black hole is likewise fixed, at the value $m\,=\,\Theta\left(L/\mathcal{A},\,0\right)$.

The angular momentum of the original black hole is $\mathcal{A}\times\mathcal{M}\,=\,\mathcal{A}\Theta\left(\mathcal{A}/L,\,\mathcal{P}\right),$ while that of the emitted object is $\left(L^2/\mathcal{A}\right)\times m\,=\, \left(L^2/\mathcal{A}\right)\Theta\left(L/\mathcal{A},\,0\right).$ We need to ensure that the difference is positive for physically acceptable values of the known parameters $\mathcal{A}$ and $\mathcal{P}$.
\begin{figure}[!h]
\centering
\includegraphics[width=0.5\textwidth]{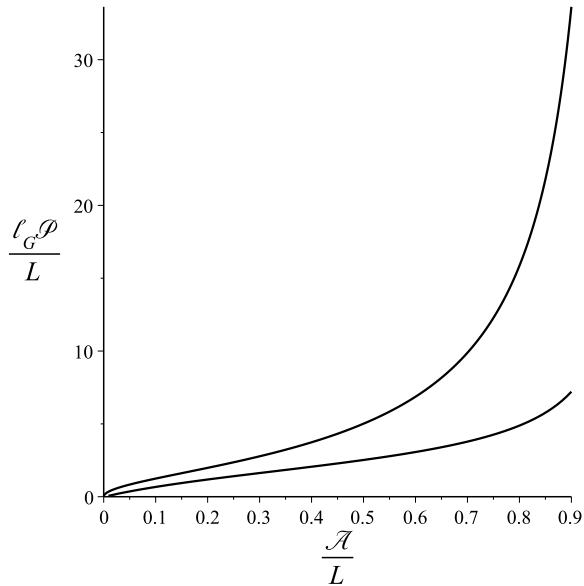}
\caption{Lower curve: $\mathcal{A}\Theta\left(\mathcal{A}/L,\,\mathcal{P}\right)\,-\,\left(L^2/\mathcal{A}\right)\Theta\left(L/\mathcal{A},\,0\right)\,=\,0$. Upper curve: $\ell_{\textsf{G}}\mathcal{P}_{\textsf{min}}/L\;=\;2\,\sqrt{\pi \mathcal{A}/L}/\left(1-{\mathcal{A}\over L}\right).$}
\end{figure}

The points below the lower curve in Figure 7 correspond to initial black holes such that angular momentum conservation forbids a final state satisfying our proposed equation (\ref{P}). However, all such black holes lie below the upper curve, which gives the minimal possible charge for the initial black hole to be stable against the emission of superradiant modes (see the inequality (\ref{H})). We see, then, that if the initial black hole is stable against emission of superradiant modes, it has sufficient angular momentum\footnote{Since $L/\mathcal{A} > 1$ and $\mathcal{A}/L < 1,$ this calculation simultaneously shows that the prospective emitted black hole has a smaller mass than the original black hole, as of course it should.} to emit a small black hole satisfying (\ref{P}).

In the course of this discussion, we saw that, if we use (\ref{P}), then the parameters of the emitted black hole are completely determined by those of the initial black hole. The same is true of the ``survivor''. Let us see exactly how that works.

The angular momentum of the survivor is  $\mathcal{A}\Theta\left(\mathcal{A}/L,\,\mathcal{P}\right)\,-\,\left(L^2/\mathcal{A}\right)\Theta\left(L/\mathcal{A},\,0\right);$ dividing this by $\mathcal{M} + \delta\mathcal{M},$ we have the angular momentum per unit mass of the survivor. Since the survivor is taken to be extremal, we have (from (\ref{F}))
\begin{equation}\label{Q}
\mathcal{M} + \delta\mathcal{M}\;=\;\Theta\left({\mathcal{A}\Theta\left(\mathcal{A}/L,\,\mathcal{P}\right)\,-\,\left(L^2/\mathcal{A}\right)\Theta\left(L/\mathcal{A},\,0\right)\over \mathcal{M} + \delta\mathcal{M}},\,\mathcal{P}\right).
\end{equation}
This equation can be solved (numerically, with explicit input data) for $\delta\mathcal{M},$ and so we have the mass of the survivor, $\mathcal{M} + \delta\mathcal{M}$. Thus indeed the parameters of the final pair are completely fixed in terms of the parameters of the original black hole.

A worked example may be helpful. Let us consider an extremal AdS$_4$-Kerr-Newman black hole with $\mathcal{P} = 100 L/\ell_{\textsf{G}}$ and $\mathcal{A}/L = \sqrt{2/3},$ the latter value chosen so that, if we use (\ref{P}), then the emitted black hole has $a/L = \sqrt{3/2};$ this value is distinguished in a way to be explained later, but any value for $a/L$ greater than unity would also work here. (From equation (\ref{H}), $\mathcal{P}_{\textsf{min}} \approx 17.456 L/\ell_{\textsf{G}}$, so superradiance is not an issue for this black hole.) Using (\ref{F}) we find that the mass is $\mathcal{M} \approx 246.26L/\ell^2_{\textsf{G}}$; this is ``large'' in the sense that it is a large multiple of the Planck mass $1/\ell_{\textsf{G}}$, since, as explained earlier, $L/\ell_{\textsf{G}}$ is normally assumed to be a large number. It is also large in the sense that, as we are about to show, the mass of the emitted black hole is much smaller. The angular momentum of this black hole is $(\sqrt{2/3})\mathcal{M}L \approx 201.07 L^2/\ell^2_{\textsf{G}}.$

The mass of the emitted black hole, found by substituting $a/L = \sqrt{3/2}$ and $p = 0$ into (\ref{F}), is $\approx 8.411L/\ell^2_{\textsf{G}}$, which is indeed much smaller than $\mathcal{M}.$ Its angular momentum is found by multiplying this mass by $\sqrt{3/2}L,$ and is $\approx 10.302 L^2/\ell^2_{\textsf{G}}.$

The angular momentum of the survivor is therefore  $\approx 190.77 L^2/\ell^2_{\textsf{G}},$ and its mass is $\approx \left(246.26 L/\ell^2_{\textsf{G}}\right)\,+\,\delta \mathcal{M}.$ Substituting all this into and solving (\ref{Q}) numerically, one finds that $\delta\mathcal{M} \approx -\,8.8232 L/\ell^2_{\textsf{G}}.$ Notice that the decrease in the mass of the original black hole is substantially larger than the mass of the emitted black hole: the latter is moving away at an initially relativistic speed, so the two objects do separate. The mass of the survivor is $\approx
237.44 L/\ell^2_{\textsf{G}}.$ The value of $\mathcal{A}/L$ drops from its original value, $\sqrt{2/3} \approx 0.8165$, to a somewhat smaller value for the survivor, $\approx 0.8034.$ This always happens when $\mathcal{A}/L < 1:$ the original black hole ``loses more angular momentum than mass''.

Thus, with equation (\ref{P}), all of the parameters of the final pair of black holes are indeed fixed by those of their progenitor, and the actual numerical values are reasonable.

\addtocounter{section}{1}
\section* {\large{\textsf{4. The Lower Bound on the Entropy Difference }}}
We can now calculate, or put bounds on, the entropies of the three black holes. We will ignore the entropy of any gravitational (or other) radiation produced in the course of the fission process; including it would of course only strengthen our argument that the final entropy exceeds the initial entropy.

For the original black hole, and for the emitted black hole, we use the equation (\ref{K}) directly. Thus we have, for the emitted black hole (using (\ref{P})),
\begin{equation}\label{R}
S\left(\textsf{Emitted}\right)\;=\;{\pi \,L^2\left(-1 + {5L^2\over \mathcal{A}^2} + \sqrt{\left(1 + {L^2\over \mathcal{A}^2}\right)^2 + {12L^2\over \mathcal{A}^2}}\right)\over 6\,\ell_{\textsf{G}}^2\left(\left(L/\mathcal{A}\right)^2 - 1\right)},
\end{equation}
and for the original black hole,
\begin{equation}\label{S}
S\left(\textsf{Original}\right)\;=\;{\pi \,L^2\left(-1 + {5\mathcal{A}^2\over L^2} + \sqrt{\left(1 + {\mathcal{A}^2\over L^2}\right)^2 + {12\over L^2}\left(\mathcal{A}^2\,+\,{\ell_{\textsf{G}}^2\left(1 - [\mathcal{A}^2/L^2]\right)^2\mathcal{P}^2\over 4\pi}\right)}\right)\over 6\,\ell_{\textsf{G}}^2\left(1 - (\mathcal{A}/L)^2\right)}.
\end{equation}

Unfortunately, it is not practical to use (\ref{K}) directly for the survivor; its entropy has to be computed by solving equation (\ref{Q}) for $\delta\mathcal{M}$ (so that all of the parameters of the black hole are known) and, while that is a straightforward numerical exercise in explicit examples, the general algebraic solution is much too complicated to be useful. Instead we make the following simple observation: among possible extremal ``survivors'' with a given charge, those with zero angular momentum are the ones with the least entropy: see Figure 3. (The Figure is drawn for $\ell_{\textsf{G}}\mathcal{P}/L = 100$, as in our worked example above, but the statement is true in general\footnote{Incidentally, it is also true that, among emitted black holes with a given value of $a/L$ ---$\,$ which, as we know, must exceed unity ---$\,$ that are extremal, the uncharged ones we are considering here are also the ones with the lowest entropies: this can be seen at once from Figure 5. (Figure 5 takes $a/L = \sqrt{3/2},$ as in our worked example, but, again, the graph has a similar form for all $a/L > 1.$)}.) The entropies of those objects therefore put a lower bound on the entropy of the survivor. This lower bound is just
\begin{equation}\label{T}
S\left(\textsf{Survivor}\right)\;\geq\;{\pi \,L^2\over 6\,\ell_{\textsf{G}}^2}\left(-1 + \sqrt{1 + {3\,\ell^2_{\textsf{G}}\mathcal{P}^2\over \pi L^2}}\right).
\end{equation}

The difference between the total entropy of the final pair and the entropy of the original black hole is therefore bounded as follows:
\begin{equation}\label{U}
S\left(\textsf{Emitted}\right)\;+\;S\left(\textsf{Survivor}\right)\;-\;S\left(\textsf{Original}\right)\;\geq \;{L^2\over \ell_{\textsf{G}}^2}\,\Psi\left(\mathcal{A},\,\mathcal{P}\right),
\end{equation}
where $\Psi\left(\mathcal{A},\,\mathcal{P}\right)$ (which is dimensionless) is defined as the sum of the right sides of relations (\ref{R}) and (\ref{T}) minus that of the right side of (\ref{S}) (with the factor $L^2/\ell_{\textsf{G}}^2$ taken out).

We now make the following crucial observation: \emph{the function $\Psi\left(\mathcal{A},\,\mathcal{P}\right)$ is strictly positive everywhere in the domain of non-negative values for $\mathcal{A}$ and $\mathcal{P}.$}

To see this, we observe first that, strictly speaking, $\Psi\left(\mathcal{A},\,\mathcal{P}\right)$ is not defined when $\mathcal{A} = 0$; but it has a well-defined limit when $\mathcal{A} \rightarrow 0:$ the limit is $\pi$ for all values of $\mathcal{P}$. Similarly, $\Psi\left(\mathcal{A},\,\mathcal{P}\right)$ is equal to $\pi$ when $\mathcal{P} = 0$, for all values of $\mathcal{A}.$ Moving away from the axes, we find that $\Psi\left(\mathcal{A},\,\mathcal{P}\right)$ is an increasing function of both variables: see for example Figure 8, which shows that it is increasing as a function of $\mathcal{A}/L$ when $\mathcal{P}$ is fixed, as in our worked example, at $100 L/\ell_{\textsf{G}}$. Similarly, in Figure 9 one sees that it is an increasing function of $\ell_{\textsf{G}}\mathcal{P}/L$ when $\mathcal{A}/L$ is fixed, as before, at $\sqrt{2/3}.$ (These examples are completely typical: the shapes are similar for all non-trivial permitted choices.)
\begin{figure}[!h]
\centering
\includegraphics[width=0.5\textwidth]{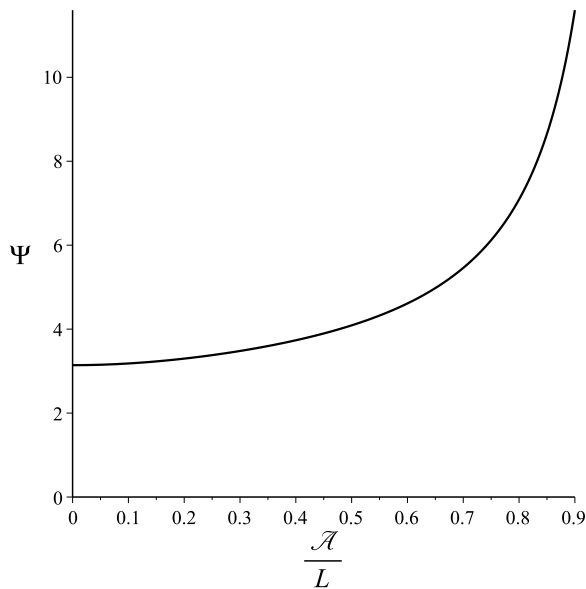}
\caption{The function $\Psi\left(\mathcal{A},\,\mathcal{P}\right)$ when $\ell_{\textsf{G}}\mathcal{P}/L = 100.$}
\end{figure}
\begin{figure}[!h]
\centering
\includegraphics[width=0.5\textwidth]{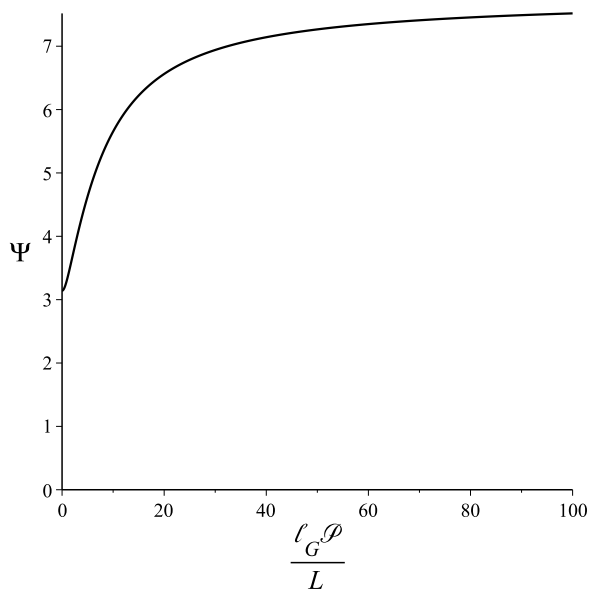}
\caption{The function $\Psi\left(\mathcal{A},\,\mathcal{P}\right)$ when $\mathcal{A}/L = \sqrt{2/3}.$}
\end{figure}

Consequently we evidently have $\Psi\left(\mathcal{A},\,\mathcal{P}\right) \geq \pi;$ and so we have finally, from the inequality (\ref{U}),
\begin{equation}\label{V}
S\left(\textsf{Emitted}\right)\;+\;S\left(\textsf{Survivor}\right)\;-\;S\left(\textsf{Original}\right)\;\geq \;{\pi L^2\over \ell_{\textsf{G}}^2}\;>\;0.
\end{equation}
That is, given any extremal AdS$_4$-Kerr-Newman black hole, it is always possible to find a pair of AdS$_4$-Kerr-(Newman) black holes which can, in accordance with the conservation laws and the second law of thermodynamics, be the result of the bifurcation of the original black hole.

This result has been obtained on the basis of Assumptions [1] and [2] discussed in Section 3, above. However, our conclusion do not depend on these assumptions being exactly correct. This is true in two senses. First, if we were to abandon Assumption [1], then values of $a/L$ smaller than $L/\mathcal{A}$, but still larger than unity, might be permitted for relatively small values of $\mathcal{P}$. But from Figure 4, this would mean that the emitted black hole has an even larger entropy than it would if (\ref{P}) holds; hence, once again, we would be underestimating the final entropy if we did so. Secondly, our eventual estimate for the final total entropy puts it well above the initial entropy; thus, even if [2] is only approximately correct, the end result is effectively the same.

To see this, consider our worked example above: the original black hole has (from equation (\ref{S})) entropy $\approx 55.091\,L^2/\ell_{\textsf{G}}^2,$ the emitted black hole has (from (\ref{R})) entropy $\approx 11.964\,L^2/\ell_{\textsf{G}}^2,$ and the survivor has (from (\ref{T})) entropy at least $\approx 54.679\,L^2/\ell_{\textsf{G}}^2,$ so that the total entropy of the final pair is at least  $\approx 66.643\,L^2/\ell_{\textsf{G}}^2,$ well in excess of the initial entropy; the difference is also well above our lower bound in (\ref{V}). We have not found any example in which the right side of (\ref{V}) does not significantly underestimate the excess of the final total entropy over the initial entropy.

The work of \cite{kn:rem1,kn:rem2,kn:goon,kn:aalsma} implies now that there exists a quantum-corrected black hole with the same mass and angular momentum as the small, ``exotic'' black hole we have considered, but with a still higher entropy. Taking all this together, we conclude that there is no thermodynamic obstruction to the bifurcation of extremal AdS$_4$-Kerr-Newman black holes to a pair consisting of a ``survivor'', together with this quantum-corrected black hole.

Let us assume, then, that extremal AdS$_4$-Kerr-Newman black holes can emit much smaller, uncharged, rotating, explicitly quantum black holes. It is difficult to say how rapidly this will occur. A definitive answer would require a full understanding of the mechanism underlying extremal bifurcation; but we can offer some simple observations.

We see from our worked example that the entropy of the emitted black hole is surprisingly large, relative to the loss of entropy by the original black hole. In the worked example, the ratio is $\approx 29.039$, and it is much larger for more massive black holes: for example, if we take an initial black hole with charge $\mathcal{P} = 500 L/\ell_{\textsf{G}},$ again with $\mathcal{A}/L = \sqrt{2/3},$ therefore with mass $\mathcal{M} \approx 2585.2L/\ell^2_{\textsf{G}},$ the ratio of the entropy of the emitted black hole to the loss of entropy by the original black hole turns out to be $\approx 294.525.$ In fact, the numerical evidence strongly suggests that the decrease in the entropy of the original black hole tends to zero with increasing mass; this means of course that, for very massive black holes, the entropy gain through bifurcation is given, to a good approximation, simply by $S\left(\textsf{Emitted}\right);$ see equation (\ref{R}) above.

Thus, by splitting, the system gains a substantial amount of entropy at very little ``cost'', and this effect is particularly marked for \emph{large} black holes. The actual numerical values of the various entropies are determined by the quantity $L/\ell_{\textsf{G}}$, which can in principle be computed with the aid of the AdS/CFT correspondence; as mentioned earlier, $L/\ell_{\textsf{G}}$ is fixed by the number of coincident M2 branes in the dual theory, which is normally assumed to be a large number. This suggests that the initial black hole corresponds to a metastable system which is easily destabilized; it may well be, therefore, that the fission takes place relatively quickly. Unlike the case of Hawking radiation, it seems likely that it occurs more readily for massive black holes than for smaller ones.

Turning these observations into a precise calculation of the fission rate is challenging, and a full treatment must await a detailed analysis of the bifurcation mechanism. It might however be possible to obtain estimates using the methods of black hole thermodynamics when all three black holes are classical, as in our calculation above; we expect this to put a lower bound on the actual rate.

\addtocounter{section}{1}
\section* {\large{\textsf{5. After the Bifurcation}}}
If we continue to assume that the ``survivor'' black hole is itself a large extremal AdS$_4$-Kerr-Newman black hole, it sheds angular momentum and mass in the same manner as its progenitor. This will continue until one of two fates overtakes an eventual descendant: either it ceases to be a classical object altogether, in which case the further development is a matter for a quantum theory of gravitation; or it becomes an AdS$_4$-Reissner-Nordstr\"{o}m black hole, which can be dealt with by means of the techniques discussed in \cite{kn:hod}. (We have assumed throughout that the emitted object has either angular momentum or charge. In reality, it might have both, so that the original black hole steadily loses both angular momentum and charge; we suppose that this case can be understood as a combination of the cases considered here and in \cite{kn:hod}.)

The fate of the \emph{emitted} black hole is much less clear. As we have stressed, it is a quantum-corrected black hole, so no firm conclusions can be drawn as yet, and such black holes are not our principal concern in this work; but let us conclude with some observations which may suggest a line of attack on the problem in the rotating case. Again we continue to assume that this black hole is extremal.

We wish to argue that, in most cases at least, the fate of the emitted black hole can be understood by embedding it in string theory, as in \cite{kn:hawrot}.

The emitted black hole still has a well-defined angular momentum to mass ratio $a$, which still satisfies $a/L > 1.$ We can assume that, like its classical counterparts, it does not experience any analogue of a superradiant instability. The novel feature arising from the embedding in string theory is that we now have to take into account the existence of \emph{branes}, objects with no classical counterpart.

The presence of branes means that the system becomes sensitive to the asymptotic spacetime geometry. Seiberg and Witten \cite{kn:seiberg} showed that, under certain general conditions, this sensitivity can cause the system to become unstable. Let us first discuss this at the classical level.

In the case of a classical asymptotically AdS black hole spacetime, the action of a brane located at a fixed value of the radial coordinate depends both on the local black hole parameters like $a$, as well as on the asymptotic curvature parameter $L$. One finds \cite{kn:104} that $a$ and $L$ have opposite effects: large values of $a$ tend to force the brane action to become unbounded below, while large values of $L$ have the opposite effect. This means that the system is stable for some values of $a/L$ but not for others. If it is not stable, the instability can be pictured as a \emph{pair-production instability} for branes \cite{kn:maldacena}.

In the AdS$_4$-Kerr-Newman case, it turns out (see \cite{kn:104} for the details) that the necessary and sufficient condition to avoid an action for branes which is unbounded below is
\begin{equation}\label{W}
a/L \;\leq\;\sqrt{3/2}\, \approx \, 1.225.
\end{equation}
Thus, in string theory, AdS$_4$-Kerr-Newman black holes with $a/L > 1$ can be completely stable, though admittedly only for a very restricted range of values for $a/L$. The corresponding minimal value for $\mathcal{A}/L$ is of course $\sqrt{2/3},$ the value we chose for our worked example above.

We should note immediately that the fact that AdS$_4$-Kerr-Newman black holes with $a/L > \sqrt{3/2}$ are unstable in string theory does not mean that they cannot exist. The ``Seiberg-Witten instability'' is non-perturbative: a barrier has to be overcome for the brane pairs to be created \cite{kn:maldacena}. Thus one should regard AdS$_4$-Kerr-Newman black holes with $a/L > \sqrt{3/2}$ as metastable objects with well-defined entropies.

A very remarkable aspect of this Seiberg-Witten instability is that the brane action is always positive \emph{near} to the black hole. The region of negative action, if it exists, is typically far from the black hole, in a region of spacetime where the curvature is relatively small, and which is therefore not sensitive to the higher-dimension operator modifications of the local black hole geometry. For example, Figure 10 shows a typical graph of the action for the region outside the event horizon, with $a/L = 1.3,$ above the critical value $\sqrt{3/2}.$ Clearly the instability will involve branes propagating in the region with $r$ exceeding about $100 L$, compared with $r_{\textsf{H}} \approx 2.242 L$ for the horizon in this example.
\begin{figure}[!h]
\centering
\includegraphics[width=0.5\textwidth]{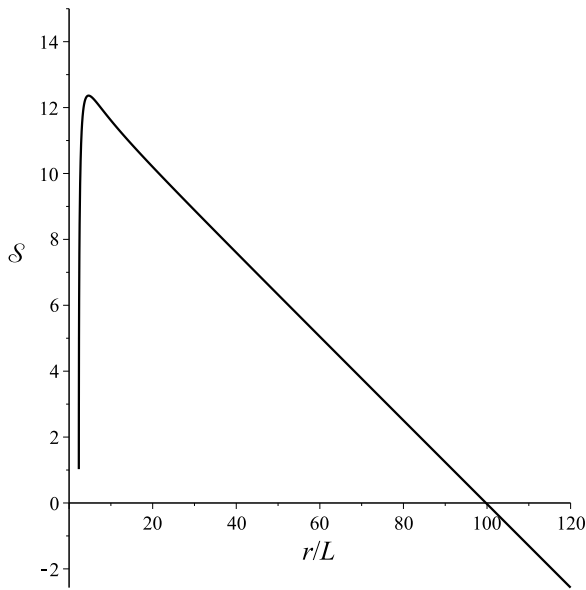}
\caption{(Scaled) action of a BPS brane outside the event horizon of an AdS$_4$-Kerr-Newman black hole with $a/L  = 1.3 \,> \sqrt{3/2}$.}
\end{figure}

We wish to argue on this basis that string theory will continue to impose a condition similar to (\ref{W}) even in the quantum-corrected case, though of course the precise numerical value on the right may differ. Therefore we can develop our argument at the qualitative level by continuing to use the classical geometry.

Let us consider, then, a classical AdS$_4$-Kerr-Newman black hole with $\mathcal{A}/L < \sqrt{2/3} \approx 0.81649;$ notice that this covers most of the available range for non-exotic black holes. From equation (\ref{P}), we see that the emitted black hole resulting from extremal bifurcation will have $a/L > \sqrt{3/2},$ so that it is subject to the Seiberg-Witten instability. It must therefore eventually shed angular momentum and mass by creating brane pairs.

One sees from Figure 2 that, if the black hole continues to be extremal, then $a/L$ must \emph{increase} in this case as the mass decreases (just as, in our worked example above, where $\mathcal{A}/L < 1,$ it had to decrease as the mass decreased). That is, the mass must decrease more rapidly than the angular momentum. But if $a/L > \sqrt{3/2}$ initially, then this means that after generating the pair, the black hole will have a still larger value of $a/L$, so again it must emit a pair of branes, and so on: $a/L$ runs away to larger and larger values until the black hole ceases to exist.

This is of course consistent with the notion that extremal black holes cannot be completely stable: the emitted black hole, too, is (assumed to be) extremal, so it too should not be able to survive indefinitely.

We are left with the relatively small set of initial black holes with $\mathcal{A}/L$ close to unity ($\sqrt{2/3} \approx 0.81649 \leq \mathcal{A}/L < 1$). Since (\ref{W}) is both necessary and sufficient, the emitted black hole in this case is stable against the Seiberg-Witten effect; nor have we been able to show that this black hole can split into a pair of black holes with a higher total entropy. 

The following observation may be relevant. Requiring that classical Cosmic Censorship should hold for these black holes, and that they should not be subject to a superradiant instability, one can show that all of them have mass at least $26.83\,L/\ell_{\textsf{G}}^2$ (see Figure 1). Since $L/\ell_{\textsf{G}}$ is always assumed to be very large, this means that the masses are very large multiples of the Planck mass, $1/\ell_{\textsf{G}}.$ This appears to contradict the assertion that these black holes are necessarily quantum-gravitational objects; in which case we can reject this class as unphysical. Perhaps this argument can also be made to work when classical Censorship is replaced by its quantum counterpart \cite{kn:kats,kn:NAH}.

\addtocounter{section}{1}
\section* {\large{\textsf{6. Conclusion}}}
Extremal black holes have attracted a great deal of theoretical attention of late, but most of the focus has been on the Reissner-Nordstr\"{o}m case. Near-extremal astrophysical black holes, however, are in that state due to rotation (though we should not ignore the exciting possibility of future observations of near-extremal magnetic black holes \cite{kn:juan1,kn:yang,kn:ullah}). Clearly, then, extremal AdS-Kerr-Newman black holes deserve more attention.

The WGC is founded on the idea that all extremal black holes, including those which are apparently classical, should decay by emitting particles or by undergoing a bifurcation (or both). We have studied the second decay channel by transferring the problem to the asymptotically AdS arena. We have found that, when one does so, a surprise awaits: it is actually easier to study extremal bifurcation in the AdS$_4$-Kerr-Newman context than in the AdS$_4$-Reissner-Nordstr\"{o}m case, in the sense that classical Cosmic Censorship need not be violated in the former situation. This permits us to evaluate explicitly the entropies of all of the black holes involved, and to show that the second law of thermodynamics mandates the splitting, if, as conjectured, a mechanism for such a process actually exists.

We have made very little use of the AdS/CFT correspondence here ---$\,$ in fact, we only mentioned it in connection with the claim that $L/\ell_{\textsf{G}}$ is a large number. Clearly there is an opportunity here to use the correspondence to shed more light on the physics of black hole bifurcation, as was attempted for example in \cite{kn:urban}, and the results of that work should be extended to the AdS$_4$-Kerr-Newman case.

On the other hand, the fact that the results of the present work do not materially depend on the global structure of the spacetime offers hope that they may be relevant, in some way, to the asymptotically flat case. For example, the fact that our results are valid for \emph{all} sufficiently large values of the asymptotic curvature parameter $L$ encourages us to hope that similar results can be proved for rapidly rotating near-extremal asymptotically flat black holes.

\addtocounter{section}{1}
\section*{\large{\textsf{Acknowledgements}}}
The author is grateful to Professor Ong Yen Chin and to Dr. Soon Wanmei for useful discussions.

\end{document}